\documentclass[a4paper]{jpconf}
\usepackage{graphicx}
\begin{document}
\title{Experiment, theory and the Casimir effect}

\author{V. M. Mostepanenko\footnote{On leave from
Noncommercial Partnership
``Scientific Instruments'',  Moscow,  Russia}}

\address{Center of Theoretical Studies and Institute for Theoretical
Physics, Leipzig University, {\protect \\}
Vor dem Hospitaltore 1, 100920,
D-04009, Leipzig, Germany}

\ead{vladimir.mostepanenko@itp.uni-leipzig.de}

\begin{abstract}
Several problems at the interface between the field-theoretical description
of the Casimir effect and experiments on measuring the Casimir force are
discussed. One of these problems is connected with the definition of
the Casimir free energy in ideal metal rectangular boxes satisfying the
general physical requirements. It is shown that the consideration of
rectangular boxes with a partition (piston) does not negate the previously
known results obtained for boxes without a piston. Both sets of results
are found to be in mutual agreement. Another problem is related to the
use of the proximity force approximation for the interpretation of the
experimental data and to the search of analytical results beyond the PFA
based on the first principles of quantum field theory. Next, we discuss
concepts of experimental precision and of the measure of agreement
between experiment and theory. The fundamental difference between these
two concepts is clarified. Finally, recent approach to the thermal
Casimir force taking screening effects into account is applied
to real metals. It is shown that this approach is thermodynamically
and experimentally inconsistent. The physical reasons of this inconsistency
are connected with the violation of thermal equilibrium which is the basic
applicability condition of the Lifshitz theory.
\end{abstract}

\section{Introduction}

For almost 50 years since Casimir's discovery [1], theory existed
independent of rare experiments [2,3]. A lot of theoretical work has been
done during this period. However, the relationship to reality of some
theoretical
models, such as an ideal metal spherical shell, an ideal metal
rectangular box, a dielectric ball etc., remains unclear up to now.
In the last ten years scientific investigations in the field of the
Casimir effect have experienced an interaction between experiment and theory.
This has revealed that the application of some basic theories to real
experimental situations is highly nontrivial and even leads to
communication difficulties between theorists and experimentallists.

In this paper we summarize some experience of the interaction between
``high theory'' and real world experimental details in the last
ten years. Different points of view are considered on such problems as
agreement between experiment and theory and applicability of some
ideal models and approximate methods in real experimental situations.
It is shown that in some cases confusion arises from an inadequate use
of terminology.

In Sec.~2 we discuss an old problem of the thermal Casimir force in
ideal metal rectangular boxes and suggest a new solution which satisfies
general physical criteria. It is shown that the case of an isolated box
is independent of a box with a partition (piston). The results
for the Casimir force obtained for each of these configurations are
in mutual agreement.

Section 3 briefly reviews the proximity force approximation including
its justification from the first principles of quantum field theory
and experimental applications. In this respect a new representation for
the Casimir energy in terms of the functional determinants and
scattering matrices is considered. The application of this
representation to real material bodies is still problematic.

In Sec.~4 we consider the problem of the reliability of experiments.
It is underlined that the experimental error is an independent characteristic
of the experimental precision which should not be confused with the
measure of agreement between experiment and theory.

Section 5 is devoted to the comparison between experiment and
theory in the measurements of the Casimir force. In this respect different
approaches to the theoretical description of the Casimir force between
real metals are compared with the most precise indirect measurement
of the Casimir pressure between two parallel plates by means of
micromechanical torsional oscillator [4,5]. Special attention is paid to
uncertainties which might be introduced in the computations due to deviations
of the tabulated optical data from the data
particular to the metallic films actually used.

In Sec.~6 we consider a recent theoretical approach to the thermal Casimir
force taking into account the screening effects and diffusion currents.
We apply this approach to the case of real metals and analyze its consistency
with the principles of thermodynamics and experimental data.
Specifically we show that for metals with perfect crystal lattices the
inclusion of screening effects results in violation of the Nernst heat
theorem. The experimental data of the experiment [4,5] exclude
this approach at a 99.9\% confidence level.

Section 7 contains our conclusions and discussion.

\section{Thermal Casimir force in ideal metal rectangular boxes}

Ideal metal rectangular boxes were first considered by Lukosz [6],
Mamayev and Trunov [7,8] and Ambj{\o}rm and Wolfram [9].
This configuration attracted much attention because it was found that
the electromagnetic Casimir force in rectangular boxes can be both
attractive and repulsive depending on the ratio of sides $a_x$, $a_y$ and
$a_z$
along the $x$, $y$ and $z$ axes. The nonrenormalized Casimir energy of the
box is equal to (for simplicity we consider the massless scalar field
with Dirichlet boundary conditions)
\begin{equation}
E_0(a_x,a_y,a_z)=\frac{\hbar}{2}\sum_{n,l,p=1}^{\infty}\omega_{nlp},
\label{eq1}
\end{equation}
\noindent
where
\begin{equation}
\omega_{nlp}=\pi c\left[\Bigl(\frac{n}{a_x}\Bigr)^2+
\Bigl(\frac{l}{a_y}\Bigr)^2+\Bigl(\frac{p}{a_z}\Bigr)^2\right]^{1/2}.
\label{eq2}
\end{equation}
\noindent
The regularization of (\ref{eq1}) can be performed, e.g., using the
Epstein zeta function or the cut-off method [10].
The latter permits to find the geometric structure of infinities contained
in (\ref{eq1}). To do so, one replaces $E_0(a_x,a_y,a_z)$ from (\ref{eq1})
with $E_0^{(\delta)}(a_x,a_y,a_z)$ by introducing the cut-off function
\begin{equation}
f(\delta\omega_{nlp})={\rm e}^{-\delta\omega_{nlp}}
\label{eq3}
\end{equation}
\noindent
under the sign of summation in (\ref{eq1}).
After the repeated application of the Abel-Plana formula [10] to
 $E_0^{(\delta)}(a_x,a_y,a_z)$ one finds that there are three different types
of divergent quantities
in the limit $\delta\to 0$, $I_1$, $I_2$ and $I_3$ of
order $\delta^{-4}$, $\delta^{-3}$ and $\delta^{-2}$, respectively.
Then, the finite, renormalized, Casimir energy can be defined as
\begin{equation}
E_0^{\rm ren}(a_x,a_y,a_z)=\lim_{\delta\to 0}\left[
E_0^{(\delta)}(a_x,a_y,a_z)-I_1-I_2-I_3\right].
\label{eq4}
\end{equation}
\noindent
Here, $I_k$ ($k=1,\,2,\,3$) are the counter terms having the following
geometrical structure:
\begin{equation}
I_1=\frac{12\pi^2\hbar a_xa_ya_z}{c^3\delta^4},\qquad
I_2=-\frac{\pi^2\hbar (a_xa_y+a_xa_z+a_ya_z)}{c^2\delta^3},\qquad
I_3=\frac{\pi\hbar (a_x+a_y+a_z)}{8c\delta^2}.
\label{eq5}
\end{equation}
\noindent
A similar situation takes place for the electromagnetic field, where
the renormalized Casimir energy, $E_{0,\rm em}^{\rm ren}$, also takes the
form of (\ref{eq4}) (with $E_0$ replaced for $E_0^{\rm em}$) and
\begin{equation}
I_1^{\rm em}=2I_1,\qquad
I_2^{\rm em}=0,\qquad
I_3^{\rm em}=-2I_3.
\label{eq6}
\end{equation}
\noindent
It is seen that in both cases the counter terms are proportional to the
volume of the box $V=a_xa_ya_z$, to the area of box surface and to the
sum of sides.

In the last few years the configuration of a rectangular box with
so-called {\it movable} partition (piston) has attracted much attention
[11--14].
This means that the piston can have any fixed position parallel to the
two opposite faces of the box (the configuration where
the piston is not fixed and
may slide between the opposite faces is in fact a nonequilibrium case).
Let the piston be parallel to the plane $xy$ and have an equation
$z=a_{z1}<a_z$. In this case our box is divided into the two boxes
$a_x\times a_y\times a_{z1}$ and $a_x\times a_y\times (a_z-a_{z1})$.
Calculating the sum of the regularized Casimir energies
\begin{equation}
E_0^{(\delta)}(a_x,a_y,a_{z1})+E_0^{(\delta)}(a_x,a_y,a_z-a_{z1}),
\label{eq7}
\end{equation}
\noindent
one finds that the contribution from the singular terms of the form of
(\ref{eq5}) does not depend on the position of the piston $a_{z1}$.
This leads to a finite force acting on the piston
\begin{equation}
F(a_x,a_y,a_z,a_{z1})=-\frac{\partial}{\partial a_{z1}}\left[
E_0^{(\delta)}(a_x,a_y,a_{z1})+E_0^{(\delta)}(a_x,a_y,a_z-a_{z1})\right].
\label{eq8}
\end{equation}
\noindent
This force is well defined and does not require the renormalization
procedure (\ref{eq4}).

In both scalar and electromagnetic cases the force acting on the piston
attracts it to the nearest face of the box. On this ground the existence
of the Casimir repulsion in cubes in the electromagnetic case was
cosidered doubtful [12]. Specifically, it was claimed [12,13] that the
definition
of the pressure acting on a cube face requires elastic deformations
of single bodies treated as perfect. The attraction (or repulsion for
a piston with Neumann boundary conditions [15]) of a piston to the nearest
face of the box does not, however, negate the Casimir repulsion for
boxes without a piston that have some appropriate ratio of $a_x$, $a_y$
and $a_z$. The point is that the cases with an empty space outside the
box and that with another section of the larger box outside the piston are
physically quite different. In the first case the vacuum energy outside the
box does not depend on $a_x$, $a_y$
and $a_z$ and there is no force acting on the box from the outside.
Whereas in the second case there is an extra section of the larger box
outside the piston which gives rise  to the additional force acting on it.
In fact one need not admit elastic deformations to define a force
and a pressure in static configurations. This is simply done using the
principle of virtual work and virtual displacements through real
forces [16,17]. In addition, from a thermodynamic point of view any
equilibrium system can be characterized by the free energy (energy if the
temperature is equal to zero) and the respective pressure [18]
\begin{equation}
P=-\left.\frac{\partial {\cal F}}{\partial V}\right|_{T={\rm const}}.
\label{eq9}
\end{equation}
\noindent
{}From this point of view it would be illogical to admit consideration
of the force acting on a piston, but exclude from consideration forces
acting on the faces of a box where this piston serves as a partition.

In this respect it seems important to provide a finite definition of the
Casimir free energy in ideal metal rectangular boxes satisfying
general physical requirements. The first calculations on this subject [9]
resulted in a divergent free energy after removing the regularization.
More recent results appear to be either infinite [19] or ambiguous [20].
Paper [21] reconsidered the derivation of the Casimir free energy in
rectangular boxes using zeta functional regularization. However, the used
formalism does not include all necessary subtractions.

The following definition of the Casimir free energy in rectangular boxes
suggests itself [12,13,21]
\begin{equation}
{\cal F}_0=E_0^{\rm ren}+\Delta_T{\cal F}_0,
\qquad
\Delta_T{\cal F}_0=k_BT\sum_{n,l,p=1}^{\infty}\ln\left(1-
{\rm e}^{-\frac{\hbar\omega_{nlp}}{k_BT}}\right).
\label{eq10}
\end{equation}
\noindent
This expression is finite. However, it cannot be considered as physically
satisfactory. The problem is that at high temperature the thermal
correction (\ref{eq10}) behaves as [22]
\begin{equation}
\Delta_T{\cal F}_0=\alpha_1\frac{(k_BT)^4}{(\hbar c)^3}+
\alpha_2\frac{(k_BT)^2}{(\hbar c)^2}+
\alpha_3\frac{(k_BT)^2}{\hbar c}+\alpha_4 k_BT+\ldots\, ,
\label{eq11}
\end{equation}
\noindent
where $\alpha_1=-V\pi^2/90$, $\alpha_{2,3}=\alpha_{2,3}(a_x,a_y,a_z)$
can be expressed in terms of the heat kernel coefficients and
$\alpha_4={\rm const}$. Then at high temperature $\Delta_T{\cal F}_0$
contains terms of quantum origin which increase with the increase
of temperature. In the general case, these terms lead to respective
forces acting on the box faces which increase with the increase of the sides
of the box. Such paradoxical properties are physically unacceptable.
Because of this it was suggested [23] to define the physical Casimir free
energy of the box as
\begin{equation}
{\cal F}=E_0^{\rm ren}+\Delta_T{\cal F}_0-
\alpha_1\frac{(k_BT)^4}{(\hbar c)^3}-
\alpha_2\frac{(k_BT)^2}{(\hbar c)^2}-
\alpha_3\frac{(k_BT)^2}{\hbar c}.
\label{eq12}
\end{equation}
\noindent
With this definition, the respective Casimir forces acting on the box
faces go to zero when all the box sides $a_x,\,a_y,\,a_z$ go to infinity
in agreement with physical intuition.

The physical meaning of all three subtractions made on the right-hand side
of (\ref{eq12}) can be clearly understood. The first term is actually
the contribution of the blackbody radiation in the volume of the box.
This is seen from the fact that the free energy density of the blackbody
radiation in empty space is given by
\begin{equation}
f_{bb}=k_BT\int\frac{d^3k}{(2\pi)^3}\ln\left(1-
{\rm e}^{-\frac{\hbar c|k|}{k_BT}}\right)=
-\frac{\pi^2(k_BT)^4}{90(\hbar c)^3},
\label{eq13}
\end{equation}
\noindent
where for the electromagnetic case $f_{bb}^{\rm em}=2f_{bb}$.

For the scalar Casimir effect in a rectangular box with sides
$a_x\times a_y\times a_z$ the asymptotic behavior of $\Delta_T{\cal F}_0$
at high $T$ was investigated in [23] with the result
\begin{equation}
\alpha_2=\frac{\zeta(3)}{4\pi}(a_xa_y+a_xa_z+a_ya_z), \qquad
\alpha_3=-\frac{\pi}{24}(a_x+a_y+a_z).
\label{eq14}
\end{equation}
\noindent
In the electromagnetic case the following values of these coefficients
were obtained:
\begin{equation}
\alpha_2^{\rm em}=0, \qquad
\alpha_3^{\rm em}=\frac{\pi}{12}(a_x+a_y+a_z).
\label{eq15}
\end{equation}
\noindent
This demonstrates that the geometric structures of all three terms
subtracted in (\ref{eq12}) are precisely the same as the terms
subtracted in (\ref{eq4}) to make the Casimir energy finite at zero
temperature. Because of this, the subtraction procedure in (\ref{eq12})
can be interpreted as the additional (finite) renormalization of the same
geometric parameters as were renormalized at zero temperature to make
the Casimir energy of the box finite.

The simplest application of the final expression for the physical
Casimir free energy (\ref{eq12}) is the case of two plane parallel plates.
It is easily seen that in this configuration $\alpha_2=\alpha_3=0$
and one is left with only a subtraction of the free energy of the
blackbody radiation in the volume between the plates $V=aS$, where $S$
is the infinite plate area. This leads to the well known result [10,24]
for the electromagnetic Casimir free energy per unit area of the plates
\begin{equation}
{\cal F}(a,T)=-\frac{\pi^2}{720a^3}
\left\{1+\frac{45}{\pi^3}\sum_{l=1}^{\infty}\left[
\frac{\coth(\pi lt)}{t^3l^3}
+\frac{\pi}{t^2l^2{\rm sinh}^2(\pi tl)}\right]-
\frac{1}{t^4}
\right\},
\label{eq16}
\end{equation}
\noindent
where $t\equiv T_{\rm eff}/T$, and the effective temperature is defined
as $k_BT_{\rm eff}=\hbar c/(2a)$. In particular, at
 $T\ll T_{\rm eff}$ one obtains
\begin{equation}
{\cal F}(a,T)=-\frac{\pi^2}{720a^3}
\left[1+\frac{45\zeta(3)}{\pi^3}
\left(\frac{T}{T_{\rm eff}}\right)^3-
\left(\frac{T}{T_{\rm eff}}\right)^4
\right],
\label{eq17}
\end{equation}
\noindent
where the last contribution on the right-hand side originates from the
subtraction of the blackbody radiation. We emphasize that only this term
contributes to the thermal correction to the electromagnetic Casimir
pressure at low temperatures (short separations)
\begin{equation}
P(a,T)=
-\frac{\pi^2}{240a^4}
\left[1+\frac{1}{3}\,
\left(\frac{T}{T_{\rm eff}}\right)^4
\right].
\label{eq18}
\end{equation}

Equation (\ref{eq12}) solves the long-standing problem on the calculation of
the physical Casimir free energies and pressures in rectangular boxes of
any size. A few examples for both the scalar and electromagnetic
Casimir effect are considered in [23]. Here we present the computational
results for the electromagnetic free energy in a cube and for the
respective Casimir force
\begin{equation}
F_x(a,T)=a^2P(a,T)=-\frac{1}{3}\,
\frac{\partial{\cal F}(a,T)}{\partial a}
\label{eq19}
\end{equation}
\noindent
acting on the opposite cube faces.

\begin{figure}[b]
\vspace*{-15.cm}
\begin{center}
\hspace*{-2.5cm}
\includegraphics{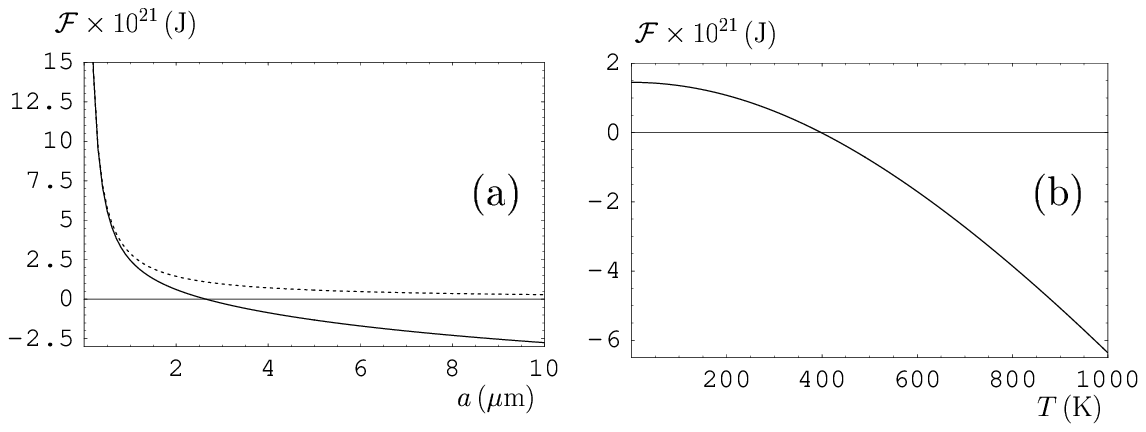}
\end{center}
\vspace*{-11.cm}
\caption{The electromagnetic Casimir free energy  for a cube as a function
of (a) size $a$ at $T=300\,$K (solid line; the dashed line
shows the energy at $T=0$) and (b) temperature at $a=2\,\mu$m.}
\end{figure}
In Fig.~1(a) we plot the electromagnetic Casimir free energy in
a cube as a function of $a$ at $T=300\,$K (solid line).
In the same figure the Casimir energy at $T=0$ is shown
by the dashed line.
As is seen in this figure, the electromagnetic Casimir free energy
decreases with the increase of separation.
At large separations
${\mathcal F}$ approaches a constant.
In Fig.~1(b) the electromagnetic Casimir free energy is shown
as a function of temperature for a cube with $a=2\,\mu$m.
The free energy decreases with the increase of $T$.
At high temperatures ${\mathcal F}$
demonstrates the classical limit.
The respective thermal electromagnetic Casimir force at $T=300\,$K,
as a function
of $a$, is shown in Fig.~2(a) by the solid line.
It is positive (i.e., repulsive) for cubes of any size.
Thus, thermal effects for cubes in the electromagnetic case increase the
strength of the Casimir repulsion.
 The dashed line in Fig.~2(a) shows the
electromagnetic Casimir force at $T=0$ as a function of $a$.
This force is given by
\begin{equation}
F_{x}(a)=\frac{0.09166}{3a^2},
\label{eq20}
\end{equation}
\noindent
i.e., it is always
repulsive.
Fig.~2(b) demonstrates the electromagnetic Casimir force
in a cube of size $a=2\,\mu$m
as a function of temperature. It is seen that the force increases
with increasing temperature.
\begin{figure}[t]
\vspace*{-15.cm}
\begin{center}
\hspace*{-2.5cm}
\includegraphics{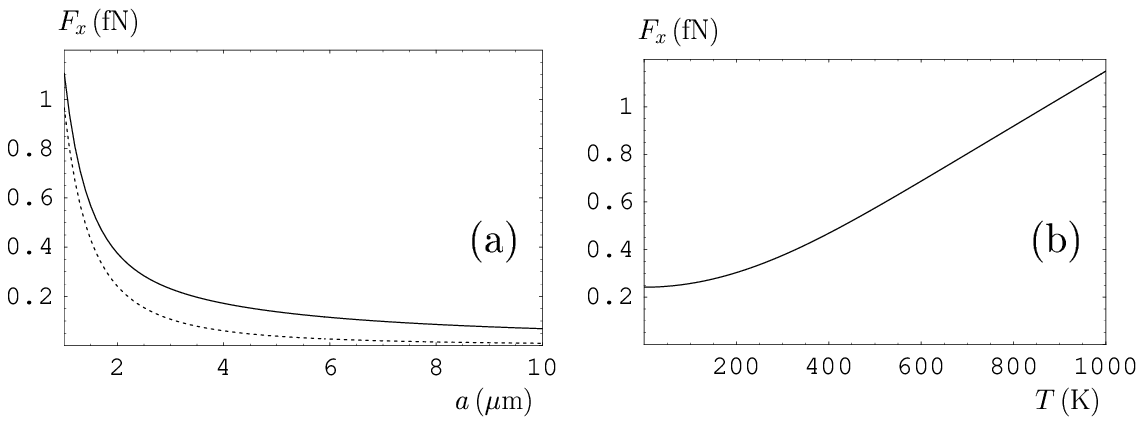}
\end{center}
\vspace*{-11.cm}
\caption{The electromagnetic Casimir force between the opposite
faces of  a cube as a function
of (a) size $a$ at $T=300\,$K (solid line; the dashed line
shows the force at $T=0$) and (b) temperature at $a=2\,\mu$m.}
\end{figure}

Note that the results presented differ from those found in [21]
where the terms of order $(k_BT)^4$ and of lower orders in the Casimir
free energy were obtained in the high-temperature regime.
 This is explained by the fact that the authors of [21]
 did not make
subtractions of the contributions from the
blackbody radiation and of the terms proportional to the box surface
area and to the sum of its sides.

The thermal correction to the Casimir
energy and force acting on a piston
were investigated in [13]
for the scalar
field with Dirichlet or Neumann boundary conditions
using the definition (\ref{eq10}).
The electromagnetic Casimir free energy and
force acting on a piston were found in the case
of ideal metal rectangular boxes and cavities with the general cross
section [13].
In the limit of low temperatures the thermal correction
to the Casimir force on a piston was shown to be exponentially small.
In the case of medium temperature $a_x\ll\hbar c/(k_BT)\ll a_y,a_z$
the authors of [13] obtained terms of order $(k_BT)^4$ and
of order $(k_BT)^2$ in the electromagnetic Casimir free energy.
In the scalar Casimir free energy, a term of order $(k_BT)^3$ was
also obtained. This results in the contribution to the force acting
on a piston which increases with the increase of the temperature, depends on
$\hbar$ and $c$ and does not depend on the position of the piston.
The scalar and electromagnetic thermal Casimir forces acting on
a piston were also considered on the basis of equation (\ref{eq10})
in [25].

The same results for the thermal correction to the Casimir force
acting on a piston are obtained if the free energy is defined in
accordance with equation (\ref{eq12}). This is because the contribution
of blackbody radiation to the energy of the entire box
is equal to
\begin{equation}
-a_xa_ya_{z1}\,f_{bb}-a_xa_y(a_z-a_{z1})f_{bb}=
-a_xa_ya_z\,f_{bb},
\label{eq21}
\end{equation}
\noindent
i.e., it does not depend on the position of the piston.
This is also true for terms of order $(k_BT)^3$ and $(k_BT)^2$
which are proportional to the surface area of each section of the box
and to the sum of its sides.

The above results were obtained for rectangular boxes with
the Dirichlet boundary conditions (scalar case) and for ideal
metal boxes (electromagnetic case). In the same way, as for zero
temperature, the consideration of the thermal Casimir effect
in rectangular boxes has to incorporate real material properties
of the boundary surfaces. Till now this problem has not been
conclusively solved.

\section{Functional determinants and the justification of the proximity
force approximation}

The proximity force approximation [26] provides an important bridge
between experiment and theory. Experimentally it is hard to use the
configuration of two parallel plates. Because of this, most of experiments
use the configuration of a sphere above a plate for which, even in the ideal
metal case, the exact results for the electromagnetic Casimir force are
not available. According to the proximity force approximation (PFA),
the interaction energy between two curved surfaces $\Sigma_1$ and
$\Sigma_2$ can be approximately calculated by replacing the small curved
surface elements with respective plane plates. If the interaction energy
between the opposite plane parallel elements is notated as $E(z)$
(where $z$ is the separation distance), the interaction energy and force
are approximately represented as
\begin{equation}
U(a)=\int_{\Sigma_1}E(z)d\sigma, \qquad
F(a)=-\frac{\partial U(a)}{\partial a}.
\label{eq22}
\end{equation}
\noindent
For the configuration of an ideal metal sphere of radius $R$ at a separation
$a$ above an ideal metal plane (\ref{eq22}) results in
\begin{equation}
F_{\rm PFA}^{s}(a)=2\pi RE(a)=-\frac{\pi^3\hbar cR}{360a^3}.
\label{eq23}
\end{equation}
\noindent
For an ideal metal cylinder above an ideal metal plate the PFA leads to
\begin{equation}
F_{\rm PFA}^{c}(a)=\frac{15\pi}{16} \sqrt{\frac{2R}{a}}E(a)=
-\frac{\pi^3}{384\sqrt{2}}\sqrt{\frac{R}{a}}\frac{\hbar c}{a^3}.
\label{eq24}
\end{equation}
\noindent
Equations (\ref{eq22})--(\ref{eq24}) are the approximate ones. They are
applicable only at short separations between the surfaces. Thus,
(\ref{eq23}) and (\ref{eq24}) work well only at $a\ll R$.

In many papers the PFA (\ref{eq22}) is applied in a region where it is
not applicable, for example at $a=R/2$. The obtained large deviations
of the PFA result from the exact result are then considered as a
``violation of the PFA''. Such formulations are in fact misleading.
The PFA gives only the main contribution to the force under some
conditions. Specifically, it would be meaningless to calculate the
integral in (\ref{eq22}) up to higher orders in the related small parameter
with the aim of obtaining a more exact result. What is really meaningful
is the search of an exact analytical representation
for the Casimir force in configurations where only the PFA result
is so far available.

In the last few years
the finite representation for the Casimir energy for two separated
bodies $A$ and $B$ in terms of the functional determinants was obtained.
In this representation the Casimir energy can be written in the form [27,28]
\begin{equation}
E(a)=\frac{1}{2\pi}\int_{0}^{\infty}\!\!\!d\xi\,
{\rm Tr}\ln\bigl(1-{\cal T}^A{\cal G}_{\xi,AB}^{(0)}
{\cal T}^B{\cal G}_{\xi,BA}^{(0)}\bigr)
=
\frac{1}{2\pi}\int_{0}^{\infty}\!\!\!d\xi\,
\ln{\rm det}\bigl(1-{\cal T}^A{\cal G}_{\xi,AB}^{(0)}
{\cal T}^B{\cal G}_{\xi,BA}^{(0)}\bigr).
\label{eq25}
\end{equation}
\noindent
Here, ${\cal G}_{\xi,AB}^{(0)}$ is the operator for the free space
Green function with the matrix elements
$\langle\mbox{\boldmath$r$}|{\cal G}_{\xi,AB}^{(0)}|
\mbox{\boldmath$r$}^{\prime}\rangle$ where
$\mbox{\boldmath$r$}$ belongs to the body $A$ and
$\mbox{\boldmath$r$}^{\prime}$ to $B$.
${\cal T}^A\,({\cal T}^B)$ is the operator of the $T$-matrix for a body
$A$ and $B$, respectively. The latter is widely used in light
scattering theory, where it is the basic object for expressing the
properties of the scatterers [29].
Using such a representation, in [28]
the analytic results for the electromagnetic Casimir energy
for an ideal metal cylinder above an ideal metal plane were obtained.
Eventually, the result is expressed through the determinant of an
infinite matrix with elements given in terms of the Bessel
functions. The analytic
asymptotic behavior of the exact Casimir energy at short
separations was found in [30]. It results in the
following expression for the Casimir force at $a\ll R$:
\begin{equation}
F^{c}(a,0)=F_{\rm PFA}^{c}(a)
\,\left[1-\frac{1}{5}\left(\frac{20}{\pi^2}-
\frac{7}{12}\right)\frac{a}{R}\right].
\label{eq26}
\end{equation}
\noindent
The PFA result (\ref{eq24}) in this case matches with the first term on the
right-hand side of (\ref{eq26}).

Equation (\ref{eq26}) is very important. It demonstrates
that the relative error of the electromagnetic Casimir force
between a cylinder and a plate calculated using the PFA
is equal to $0.2886\,a/R$. Thus, for typical
experimental parameters of
$R=100\,\mu$m and $a=100\,$nm this error is approximately
equal to only 0.03\%.

For a sphere above a plate made of ideal metals
the exact analytic solution in the
electromagnetic case has not yet been obtained. The
scalar Casimir energy for a sphere above a plate was found in [30,31].
The scalar Casimir energies
for both a sphere and a cylinder above a plate have also been
computed numerically using the wordline algorithms [32,33],
 but it was noted that
the Casimir energies for the Dirichlet scalar field should
not be taken as an estimate for those in the electromagnetic
case.
For an ideal metal sphere above an ideal metal plane a
correction of order $a/R$ beyond the PFA was computed numerically
in [34] for $a/R\geq 0.075$ and in [35]
for $a/R\geq 0.15$. In both cases the extrapolation of the
obtained results to smaller $a/R$ leads to a coefficient
near $a/R$ approximately equal to 1.4.

In addition, the validity of the PFA for a sphere above a plate
has been estimated experimentally [36] and
the error introduced from the use of this
approximation was shown to be less than $a/R$.
 This is in disagreement with
the extrapolations made in [34,35].
To solve this contradiction, it is desirable to find
the analytical form of the first correction beyond the PFA
for a sphere above a plane,
like in (\ref{eq26}) for the cylinder-plane configuration.

In fact the representation (\ref{eq25}) provides a far-reaching
generalization of the Lifshitz formula. {}From conceptual point of view
it can be applied not only to ideal metals, but to real materials as
well. The problem, however, is to find the matrix elements of the
$T$-matrix operator which would take proper account of both geometric
shape and material properties of the test bodies used in the
experimental situation.

\section{The experimental error and reliability of experiments}

The concept of the experimental error is often confused with the
theoretical error and with the measure of agreement between experiment
and theory. However, when we deal with an {\it independent measurement},
the experimental error has nothing to do with any theory of the
measured quantity.
The independent measurement of the Casimir force or its gradient does
not use any theory of the Casimir effect. Thus, the experiments [4,5,37--42]
are independent in this respect. In other experiments (in [43], for instance)
the measurement data are fitted to some theoretical expression for the
Casimir force. Such kind of measurements are not independent
and we do not consider them below.

Some papers arrive to theoretical conclusions which are inconsistent
with the measurement data. This is sometimes surrounded by the
statement that the measurements might be not as precise as indicated by the
authors. It is our opinion that such statements made without
an indication of any specific cause are inappropriate.
Both random, $\Delta^{\!\rm rand}F^{\rm expt}(a)$, and systematic,
$\Delta^{\!\rm syst}F^{\rm expt}(a)$, experimental errors in the Casimir
force measurements are found using the rigorous statistical procedures.
They can be combined to find the total experimental error
\begin{equation}
\Delta^{\!\rm tot}F^{\rm expt}(a)=q_{\beta}(r)\left[
\Delta^{\!\rm rand}F^{\rm expt}(a)+
\Delta^{\!\rm syst}F^{\rm expt}(a)\right].
\label{eq27}
\end{equation}
\noindent
Here, $q_{\beta}(r)$ determined at $\beta=0.95$ (i.e., at 95\% confidence
level) varies between 0.71 and 0.81 depending of the value of the
quantity $r=\Delta^{\!\rm syst}F^{\rm expt}(a)/s_{\bar{F}}(a)$,
where $s_{\bar{F}}(a)$ is the variance of the mean measured
quantity [44]. In fact there is no
arbitrariness in the determination of the total experimental error which
is the ultimate characteristic of the precision of the measurements.
The most valuable esperiments are marked by a negligible role of the
random error. For such experiments
\begin{equation}
\Delta^{\!\rm tot}F^{\rm expt}(a)\approx
\Delta^{\!\rm syst}F^{\rm expt}(a).
\label{eq28}
\end{equation}
\noindent
For today there is only one indirect measurement of the Casimir pressure
between Au coated plates by means of micromechanical torsional
oscillator satisfying this condition [4,5]. The total experimental
error in this measurement at shortest separations is as small as 0.2\% of the
measured Casimir pressure. We stress once again that this error is
unrelated to much larger errors inherent to theoretical computations
on the basis of the Lifshitz theory or to the measure of agreement
between experiment and theory. This is just the resulting error with
which the experimental data are taken.
\begin{figure}[t]
\vspace*{-11.5cm}
\begin{center}
\hspace*{-2.5cm}
\includegraphics{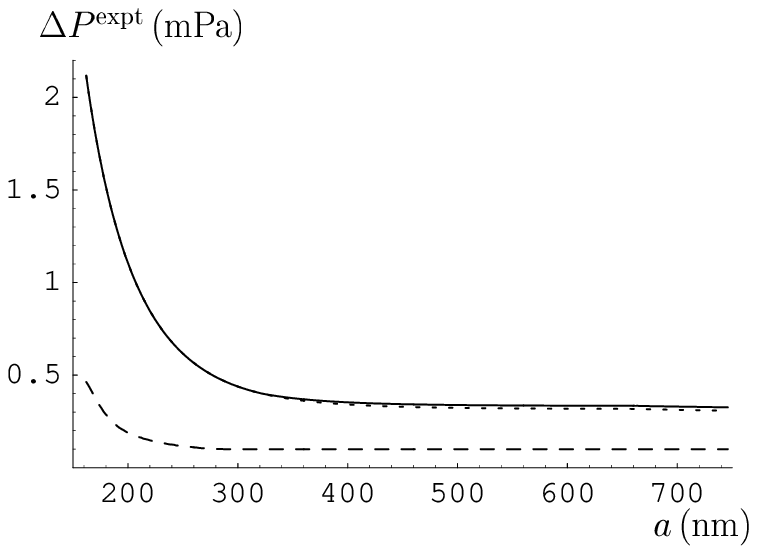}
\end{center}
\vspace*{-14.5cm}
\caption{The total absolute experimental error of the Casimir pressure
measurements [4,5] (the solid line), the random error (the long-dashed line),
and the systematic error (the short-dashed line) are
shown as functions of separation.}
\end{figure}
As an example, the total absolute experimental error in the experiment
on measuring the Casimir pressure by means of a micromechanical torsional
oscillator [4,5] is shown in Fig.~3 as a function of separation
(the solid line).
The long-dashed and short-dashed lines show the random and systematic errors,
respectively. As a result, the relative total experimental error
$\delta^{\rm tot}P^{\rm expt}(a)=
\Delta^{\!\rm tot}P^{\rm expt}(a)/|P^{\rm expt}(a)|$ varies from
0.19\% at $a=162\,$nm to 0.9\% at $a=400\,$nm, and
to 9.0\% at $a=746\,$nm.

Sometimes the experimental precision can be questioned if there are some
doubts in the calibration procedures used. For example, the electrostatic
calibration is of prime importance in the independent measurements of
the Casimir force. Specifically, it is usually carefully verified that the
residual potential between the grounded test bodies does not depend
on separation where the measurements of the electric force are performed.
Recently it was claimed that the residual potential $V_0$ from the
electrostatic calibration in the sphere-plate configuration is separation
dependent [45]. The authors used an Au-coated sphere
of 30.9\,mm radius above an Au coated plate. On the basis of these
measurements a reanalysis of the independence of $V_0$ on separation
in the earlier measurements of the Casimir force by means of an atomic force
microscope and a micromachined oscillator  was invited.
The results [45] are, however,
not directly relevant to the earlier measurements.
The point is that the radius of the sphere
used in [45] is a factor of 300 larger than in the earlier
precision measurements of the Casimir force.
It is well known that
for large test bodies (i.e., large interaction areas)
there are large variations of electric forces
due to deviations of the mechanically polished and ground lens surface
from perfect spherical shape [46].

\section{Comparison between experiment and theory}

Experiment is the supreme arbiter in physics. Because of this, the comparison
between experiment and theory is a painful point for those theories that
are found to be experimentally inconsistent. It happens that in such
cases both the experimental data and the methods of comparison are
questioned. The Casimir force is a strongly nonlinear function of the
separation distance. As a consequence, such global characteristics of
the agreement between experiment and theory as the root-mean-square
deviation were found to be inadequate [47].
In the last few years two local methods on how to compare experiment
with theory in the Casimir force measurements were elaborated and
successfully applied. Within the first method [4,48,49],
the experimental data are represented as crosses with arms determined
by the total experimental errors in the measurement of separation and
a related quantity (the force, the pressure or
the frequency shift) determined at some
chosen confidence level.
In the same figure, one should plot the theoretical band whose width
is equal to the total theoretical error determined at the same confidence
as the experimental errors. The overlap (or its absence) of the
experimental crosses and the theoretical band can be used to make
a conclusion on the consistency or inconsistency between experiment and
theory.

\begin{figure}[t]
\vspace*{-10.5cm}
\begin{center}
\hspace*{-2.5cm}
\includegraphics{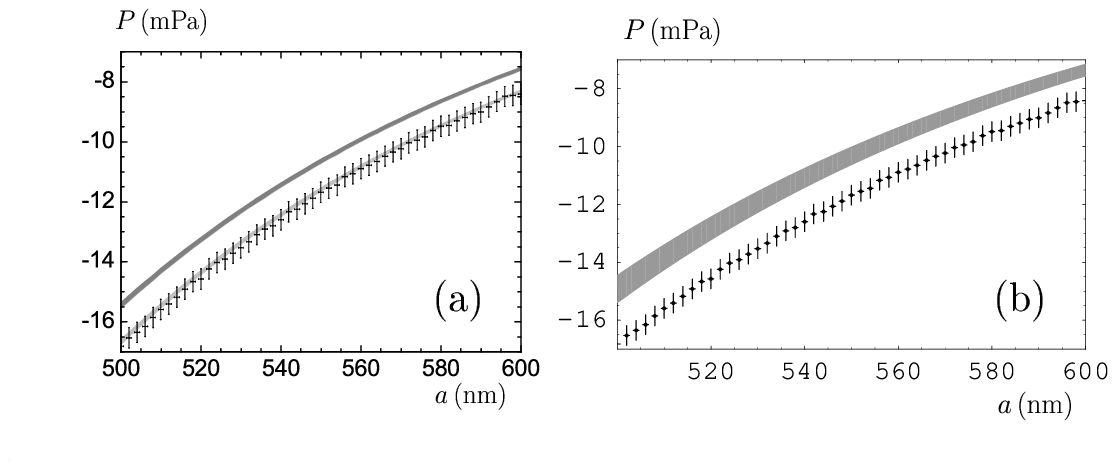}
\end{center}
\vspace*{-16.8cm}
\caption{The crosses show the measured mean Casimir pressures together
with the absolute errors
in the separation and pressure as a function of the separation.
(a) The theoretical Casimir pressures computed using
the generalized plasma-like model and the optical data extrapolated by
the Drude model are shown by the light-gray and dark-gray bands,
respectively. (b) The theoretical Casimir pressures
computed using different sets
of optical data available in the literature versus separation are shown
as the dark-gray band.}
\end{figure}
In Fig.~4 the first method of comparison between experiment and
theory is illustrated on the measurement data by Decca et al. [4,5]
discussed in Sec.~3. The light-gray band in Fig.~4(a) shows the
theoretical results computed using the Lifshitz theory combined with
the generalized plasma-like dielectric permittivity [50,51]
\begin{equation}
\varepsilon_{gp}({\rm i}\xi)=\varepsilon({\rm i}\xi)+
\frac{\omega_p^2}{\xi^2}, \qquad
\varepsilon({\rm i}\xi)=1+\sum_{j=1}^{K}
\frac{f_j}{\omega_j^2+\xi^2+\gamma_j\xi}.
\label{eq29}
\end{equation}
\noindent
Here, $\omega_p$ is the plasma frequency, $\omega_j\neq 0$ are the
frequencies of the oscillators describing  core electrons, $f_j$ are the
oscillator strengths and $\gamma_j$ are the relaxation parameters.
The dark-gray band in Fig.~4(a) is computed by the same Lifshitz theory
using the tabulated optical data for Au [52] extrapolated  to low
frequencies by means of the Drude model [53--55]
\begin{equation}
\varepsilon_{D}({\rm i}\xi)=1+
\frac{\omega_p^2}{\xi(\xi+\gamma)}=
1+\frac{4\pi\sigma({\rm i}\xi)}{\xi}.
\label{eq30}
\end{equation}
\noindent
Here $\sigma({\rm i}\xi)$ is the conductivity. It is connected
with the dc conductivity by the equation
\begin{equation}
\sigma({\rm i}\xi)=\frac{\sigma(0)}{1+\frac{\xi}{\gamma}}.
\label{eq30a}
\end{equation}
\noindent
Note that the plasma frequency and the dc conductivity are expressed as [56]
\begin{equation}
\omega_p^2=\frac{4\pi e^2n}{m}, \qquad \sigma(0)=\mu\,|e|\,n,
\label{eq30b}
\end{equation}
\noindent
where $e$ and $m$ are the charge and the mass of an electron,
$n$  is the charge carrier density and $\mu$ is their mobility.
As is seen in Fig.~4(a), the experimental data shown as crosses
(the experimental errors are determined at a 95\% confidence level)
are consistent with the theoretical approach using the
generalized plasma-like permittivity.
The Drude model approach is excluded at a 95\% confidence level.
In Fig.~4(b) the same experimental data are reproduced and compared with
the Drude model approach using all sets of optical data available
in the literature [57]. As is seen in Fig.~4(b), the use of optical
data alternative to [52] makes the disagreement deeper between the
experimental data and the Drude model approach. In Fig.~4(a,b) the comparison
between experiment and theory is performed within the separation region
from 500 to 600\,nm. However, exactly the same conclusions follow over
the entire measurement range in this experiment from 160 to 750\,nm.

In the second method for the comparison of experiment and theory in the
Casimir force measurements [40,48,58], the differences between the theoretical
and mean experimental quantity, for instance,
$P^{\rm theor}(a)-\bar{P}^{\rm expt}(a)$, are plotted as dots.
In the same figure the borders of the confidence intervals
$[-\Xi_{P}(a),\Xi_{P}(a)]$ for this difference at a chosen confidence
level (usually 95\%) are plotted as the function of separation.
If no less than 95\% of the dots representing the above differences belong
to the confidence interval the theoretical approach is consistent with
the data.
Alternatively, if almost all the dots are outside the confidence interval,
the theoretical approach is excluded by the data at a 95\% confidence level.
\begin{figure}[b]
\vspace*{-4.cm}
\begin{center}
\hspace*{-2.cm}
\includegraphics{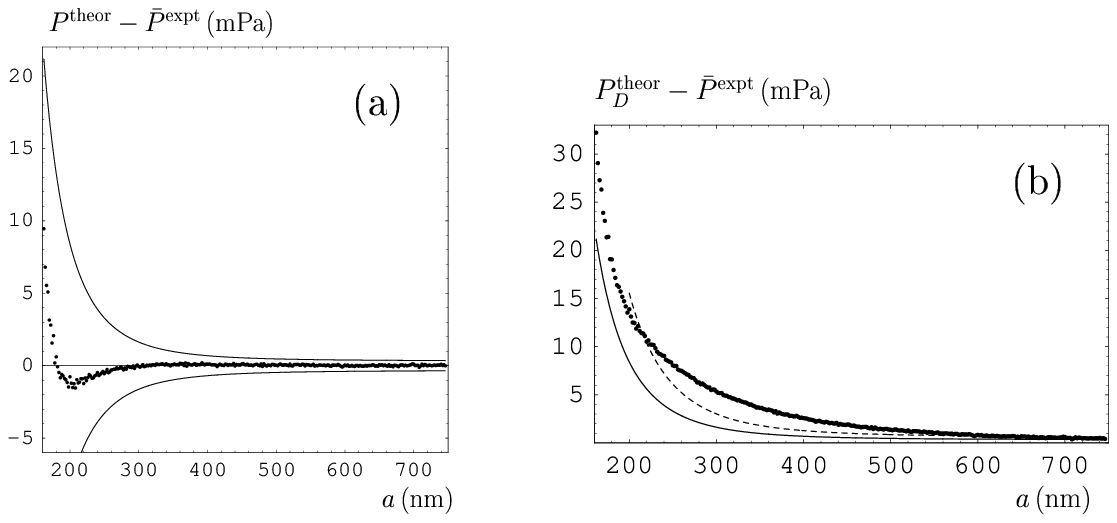}
\end{center}
\vspace*{-21.cm}
\caption{The differences of the
theoretical and the  mean experimental
Casimir pressures between the two Au plates
versus separation are shown as dots.
The theoretical results are calculated using the
Lifshitz theory at room temparature using (a) the generalized plasma-like
model and (b) the Drude model approach.
The solid lines indicate the boundaries
of the 95\% confidence intervals. The dashed line indicates the
boundary of the 99.9\% confidence intervals.
}
\end{figure}
In Fig.~5 we illustrate the second method for the comparison of
experiemnt with theory using the experimental data of the same
measurements [4,5]. In Fig.~5(a) the theoretical approach using the
generalized plasma-like permittivity (\ref{eq29}) is compared with the
data. It is seen that all dots are inside the error bars. Thus, this
approach is consistent with the data. In Fig.~5(b) the same data are
compared with the theoretical approach using the tabulated optical data
extrapolated by the Drude model. The solid and dashed line represent
the borders of 95\% and 99.9\% confidence intervals, respectively.
As is seen in Fig.~5(b), the Drude model approach is experimentally
excluded at a 95\% confidence level within the entire measurement range
from 160 to 750\,nm. Within a more narrow measurement range from 210 to
620\,nm the Drude model approach is excluded at a 99.9\% confidence
level.

If the theoretical approach is experimentally consistent [see Fig.~5(a)]
the quantity $\Xi_{P}/|\bar{P}^{\rm expt}|$, determined at a 95\%
confidence level, can be used as the quantitative measure of agreement
between experiment and theory. Thus, at $a=162\,$nm this measure is
equal to 1.9\%. It decreases to 1.4\% at $a=300\,$nm and than
gradually increases up to 9.7\% at $a=745\,$nm. It is evident that at the
shortest separation the agreement between experiment and theory is almost
an order of magnitude worse than the total experimental error equal to
only 0.19\%. This is explained by large theoretical errors which
dominate in the determination of $\Xi_{P}$ at the shortest separations.

The above explanations aim to make absolutely clear that the
calculation of errors and the comparison between experiment and theory
is not an arbitrary, but a rigorously determined procedure. Recently,
the measurement data of the experiment [4,5] was independently reanalyzed
in [59] with the conclusion: ``The data rule out the Drude approach$\ldots\,$,
while they are consistent with the plasma-model approach$\ldots$''

\section{Attempt to account for screening effects}

Recently, the above discussed problems of the Drude model approach in
application to real metals, and related problems arising for dielectric
and semiconductor materials [60--64], motivated an attempt to modify
the reflection coefficients in the Lifshitz formula by including the
screening effects and diffusion currents [65,66]. The modified reflection
coefficients for the transverse magnetic and transverse electric modes
were obtained through use of Boltzmann transport equation which takes into
account not only the standard drift current $\mbox{\boldmath$j$}$, but
also the diffusion current $eD\nabla{n}$, where $D$ is the diffusion
coefficient and $\nabla{n}$ is the gradient of the charge carrier density [66].
The transverse magnetic coefficient takes the form
\begin{equation}
\tilde{r}_{\rm TM}({\rm i}\xi,k_{\bot})=
\frac{\tilde\varepsilon({\rm i}\xi)q-k-\frac{k_{\bot}^2}{\eta({\rm i}\xi)}\,
\frac{\tilde\varepsilon({\rm i}\xi)-
\varepsilon({\rm i}\xi)}{\varepsilon({\rm i}\xi)}}{\tilde\varepsilon
({\rm i}\xi)q
+k+\frac{k_{\bot}^2}{\eta({\rm i}\xi)}\,
\frac{\tilde\varepsilon({\rm i}\xi)-
\varepsilon({\rm i}\xi)}{\varepsilon({\rm i}\xi)}},
\label{eq31}
\end{equation}
\noindent
where $k_{\bot}$ is the projection of the wave vector in the plane of
the plates, $\omega={\rm i}\xi$ is the imaginary frequency and the following
notations are introduced
\begin{eqnarray}
&&
q^2=k_{\bot}^2+\frac{\xi^2}{c^2}, \qquad
k^2=k_{\bot}^2+\tilde\varepsilon({\rm i}\xi)\frac{\xi^2}{c^2},
\qquad
\tilde\varepsilon({\rm i}\xi)=\varepsilon({\rm i}\xi)+
\frac{\omega_p^2}{\xi(\xi+\gamma)},
\nonumber \\
&&
\eta({\rm i}\xi)=\left[k_{\bot}^2+\kappa^2
\frac{\varepsilon(0)}{\varepsilon({\rm i}\xi)}\,
\frac{\tilde\varepsilon({\rm i}\xi)}{\tilde\varepsilon({\rm i}\xi)-
\varepsilon({\rm i}\xi)}\right]^{1/2}.
\label{eq32}
\end{eqnarray}
\noindent
In this equation, $1/\kappa$ is the screening length and the dielectric
permittivity
of core electrons $\varepsilon({\rm i}\xi)$ is defined in (\ref{eq29}).
The transverse electric coefficient is given by the standard expression
\begin{equation}
\tilde{r}_{\rm TE}({\rm i}\xi,k_{\bot})=
\frac{q-k}{q+k},
\label{eq33}
\end{equation}
\noindent
as is used in the Drude model approach.

The paper [66] claims the application of the
above approach to
intrinsic semiconductors only. It uses a specific Debye-H\"{u}ckel expression
for the screening length
\begin{equation}
\frac{1}{\kappa}=\frac{1}{\kappa_{\rm DH}}=R_{\rm DH}=
\sqrt{\frac{\varepsilon(0) k_BT}{4\pi e^2n}}.
\label{eq34}
\end{equation}
\noindent
This expression is applicable to particles obeying the Maxwell-Boltzmann
statistics. It is obtained from the general representation for the
screening length [67]
\begin{equation}
\frac{1}{\kappa}=R=
\sqrt{\frac{\varepsilon(0) D}{4\pi \sigma(0)}}
\label{eq35}
\end{equation}
\noindent
if one uses the expression (\ref{eq30b}) for the dc conductivity and
Einstein's relation [56,67]
\begin{equation}
\frac{D}{\mu}=\frac{k_BT}{|e|}
\label{eq36}
\end{equation}
\noindent
valid in the case of Maxwell-Boltzmann statistics.
In the limiting case $\xi\to 0$ the reflection coefficient (\ref{eq31})
coincides with that obtained in [65].

However, the application region of the reflection coefficients (\ref{eq31}),
(\ref{eq32}) with the Debye-H\"{u}ckel screening length (\ref{eq34})
cannot be restricted to only intrinsic semiconductors. These coefficients
should be applicable to all materials where the density of charge carriers
is not too large so that they are described by Maxwell-Boltzmann
statistics. This means that in the framework of the proposed approach it is
legal to apply (\ref{eq31})--(\ref{eq34}) to doped semiconductors with
dopant concentration below critical and to solids with ionic conductivity etc.

Here, we consider the application of this approach to metallic plates.
Metals and semiconductors of metallic type are characterized by rather high
concentration of charge carriers which obey the quantum Fermi-Dirac
statistics.  The general transport equation, however, is equally applicable
to classical and quantum systems. The only difference one should take into
account is the type of statistics. Substituting Einstein's relation,
valid in the case of Fermi-Dirac statistics [56,67]
\begin{equation}
\frac{D}{\mu}=\frac{2E_F}{3|e|},
\label{eq37}
\end{equation}
\noindent
where $E_F=\hbar\omega_p$ is the Fermi energy, into (\ref{eq35}),
one arrives to the following expression for the Thomas-Fermi screening
length [67]
\begin{equation}
\frac{1}{\kappa}=\frac{1}{\kappa_{\rm TF}}=R_{\rm TF}=
\sqrt{\frac{\varepsilon(0) E_F}{6\pi e^2n}}.
\label{eq38}
\end{equation}
\noindent
With this definition of the parameter $\kappa$, it is legal to apply
equations (\ref{eq31})--(\ref{eq33}) to metals.

Now we consider two thick metallic plates separated by a distance $a$
at temperature $T$ in thermal equilibrium. Under these conditions the
Casimir free energy per unit area of the plates is given by the Lifshitz
formula  [68]. Let us assume that the reflection coefficients
(\ref{eq31})--(\ref{eq33}), (\ref{eq38}) can be substituted into this
formula. Then in terms of dimensionless variables $y=2aq$,
$\zeta=\xi/\omega_c\equiv 2a\xi/c$ one obtains
\begin{equation}
\tilde{\cal F}(a,T)=\frac{k_BT}{8\pi a^2}
\sum_{l=0}^{\infty}{\vphantom{\sum}}^{\prime}
\int_{\zeta_l}^{\infty}\!\!\!y\,dy\left\{\ln\left[1-
\tilde{r}_{\rm TM}^2({\rm i}\zeta_l,y)\,{\rm e}^{-y}\right]+
\ln\left[1-
\tilde{r}_{\rm TE}^2({\rm i}\zeta_l,y)\,{\rm e}^{-y}\right]\right\},
\label{eq39}
\end{equation}
\noindent
where $\zeta_l=4\pi ak_BTl/(\hbar c)$ are the dimensionless Matsubara
frequencies and a prime near the summation sign adds a multiple 1/2 to the
term with $l=0$.
In terms of the dimensionless variables the reflection coefficient
(\ref{eq31}) takes the form
\begin{equation}
\tilde{r}_{\rm TM}({\rm i}\zeta,y)=\frac{\tilde\varepsilon y
-\bigl[y^2+(\tilde\varepsilon-1)\zeta^2\bigr]^{1/2}-
\frac{(y^2-\zeta^2)(\tilde\varepsilon-
\varepsilon)}{\tilde\eta\,
\varepsilon}}{\tilde\varepsilon y
+\bigl[y^2+(\tilde\varepsilon-1)\zeta^2\bigr]^{1/2}+
\frac{(y^2-\zeta^2)(\tilde\varepsilon-
\varepsilon)}{\tilde\eta\,
\varepsilon}},
\label{eq40}
\end{equation}
\noindent
where
\begin{equation}
\tilde\eta=2a\eta=\left[y^2-\zeta^2+\kappa_a^2\frac{\varepsilon(0)
\tilde\varepsilon}{\varepsilon
(\tilde\varepsilon-\varepsilon)}
\right]^{1/2}, \qquad
\kappa_a\equiv 2a\kappa_{\rm TF}.
\label{eq41}
\end{equation}
\noindent
Note that all dielectric permittivities here are functions of
${\rm i}\omega_c\zeta$.
Below we do not use the explicit expression for the reflection
coefficient $\tilde{r}_{\rm TE}({\rm i}\zeta,y)$ because it coincides
with the standard one, as defined in the Drude model approach, and
considered in detail in [69].

Let us determine the behavior of the Casimir free energy (\ref{eq39}) at
low temperature. For all metals the screening length (\ref{eq38}) is very
small. As a result, at any reasonable separation distance between the
plates, the dimensionless parameter $\kappa_a$ defined in (\ref{eq41})
is very large and the inverse quantity can be used as a small parameter
\begin{equation}
2a\kappa_{\rm TF}=\kappa_a\gg 1, \qquad
\beta_a\equiv\frac{1}{\kappa_a}\ll 1.
\label{eq43}
\end{equation}
\noindent
Expanding the reflection coefficient (\ref{eq40}) up to the first power
of the parameter $\beta_a$ one obtains
\begin{eqnarray}
&&
\tilde{r}_{\rm TM}({\rm i}\zeta,y)={r}_{\rm TM}({\rm i}\zeta,y)-
2\beta_a\,Z+O(\beta_a^2),
\label{eq44} \\
&&
Z\equiv\sqrt{\frac{\tilde\varepsilon
(\tilde\varepsilon-\varepsilon)^3}{\varepsilon(0)\varepsilon}}
\,\frac{y(y^2-\zeta^2)}{[\tilde\varepsilon y+\sqrt{y^2+
(\tilde\varepsilon -1)\zeta^2}]^2},
\nonumber
\end{eqnarray}
\noindent
where ${r}_{\rm TM}({\rm i}\zeta,y)$ is the standard TM reflection
coefficient calculated with the dielectric permittivity
$\tilde\varepsilon({\rm i}\omega_c\zeta)$ [it is given by (\ref{eq40})
with the third term in both numerator and denominator omitted].
{}From (\ref{eq44}) one arrives at
\begin{equation}
\ln\left[1-\tilde{r}_{\rm TM}^2({\rm i}\zeta,y)\,{\rm e}^{-y}\right]=
\ln\left[1-{r}_{\rm TM}^2({\rm i}\zeta,y)\,{\rm e}^{-y}\right]+
4\beta_a\frac{{r}_{\rm TM}({\rm i}\zeta,y)\,Z}{{\rm e}^{y}-
{r}_{\rm TM}^2({\rm i}\zeta,y)}+O(\beta_a^2).
\label{eq45}
\end{equation}

Now we substitute (\ref{eq45}) and the respective known expression for the
TE contribution [69] into (\ref{eq39}). Calculating the sum with the help
of the Abel-Plana formula, we obtain in perfect analogy to [69]
\begin{equation}
\tilde{\cal F}(a,T)={\cal F}_{gp}(a,T)-\frac{k_BT}{16\pi a^2}
\int_{0}^{\infty}\!\!\!y\,dy\,\ln\left[1-r_{{\rm TE},gp}^2(0,y)\,
{\rm e}^{-y}\right]+{\cal F}^{(\gamma)}(a,T)+
\beta_a{\cal F}^{(\beta)}(a,T),
\label{eq46}
\end{equation}
\noindent
where ${\cal F}^{(\gamma)}(a,T)$ is determined by equation (17) in [69].
It goes to zero together with its derivative with respect to temperature
when $T\to 0$. The quantity ${\cal F}^{(\beta)}(a,T)$ originates from
the second contribution on the right-hand side of (\ref{eq44}).
It is easily seen that
${\cal F}^{(\beta)}(a,T)=E^{(\beta)}(a)+O(T^3/T_{\rm eff}^3)$ at low $T$.
The Casimir free energy ${\cal F}_{gp}(a,T)$ is defined by substituting the
dielectric permittivity (\ref{eq29}) of the generalized plasma-like model
into the Lifshitz formula. It was found in [50,51] and the
respective thermal
correction was shown to be of order $(T/T_{\rm eff})^3$ when $T\to 0$.
The TE reflection coefficient at zero frequency entering (\ref{eq46})
is given by
\begin{equation}
r_{{\rm TE},gp}(0,y)=\frac{cy-\sqrt{4a^2\omega_p^2+c^2y^2}}{cy
+\sqrt{4a^2\omega_p^2+c^2y^2}}.
\label{eq47}
\end{equation}

As a result, calculating the Casimir entropy
\begin{equation}
\tilde{S}(a,T)=-\frac{\partial\tilde{\cal F}(a,T)}{\partial T}
\label{eq48}
\end{equation}
\noindent
with the use of (\ref{eq46}) and considering the limiting case of zero
temperature, one arrives at
\begin{equation}
\tilde{S}(a,0)=\frac{k_B}{16\pi a^2}\int_{0}^{\infty}\!\!\!y\,dy\,\ln
\left[1-\left(\frac{cy-\sqrt{4a^2\omega_p^2+c^2y^2}}{cy
+\sqrt{4a^2\omega_p^2+c^2y^2}}\right)^2\,{\rm e}^{-y}\right]<0
\label{eq49}
\end{equation}
\noindent
in violation of the Nernst heat theorem. This result is obtained for
metals with perfect crystal lattices. In the presence of impurities the
Casimir entropy abruptly jumps to zero at $T<10^{-3}\,$K [70].

Thus, the modified reflection coefficients taking the screening effects
into account lead to a violation of the Nernst heat theorem for metals
with perfect crystal lattices in the same way as the standard Drude model
approach. Because of this, the theoretical approach using such reflection
coefficients is thermodynamically inconsistent.

Now we briefly compare the theoretical predictions, following from the
use of reflection coefficients $\tilde{r}_{\rm TM}$ and
$\tilde{r}_{\rm TE}$, with the measurement data of the most precise experiment
by means of micromachined torsional oscillator [4,5]. This experiment
was already discussed in Sec.~5.
In Fig.~6(a) the experimental data  for the
Casimir pressure between two Au plates
are shown as crosses with the absolute errors determined
at a 95\% confidence level.
\begin{figure}[b]
\vspace*{-4.cm}
\begin{center}
\hspace*{-2.cm}
\includegraphics{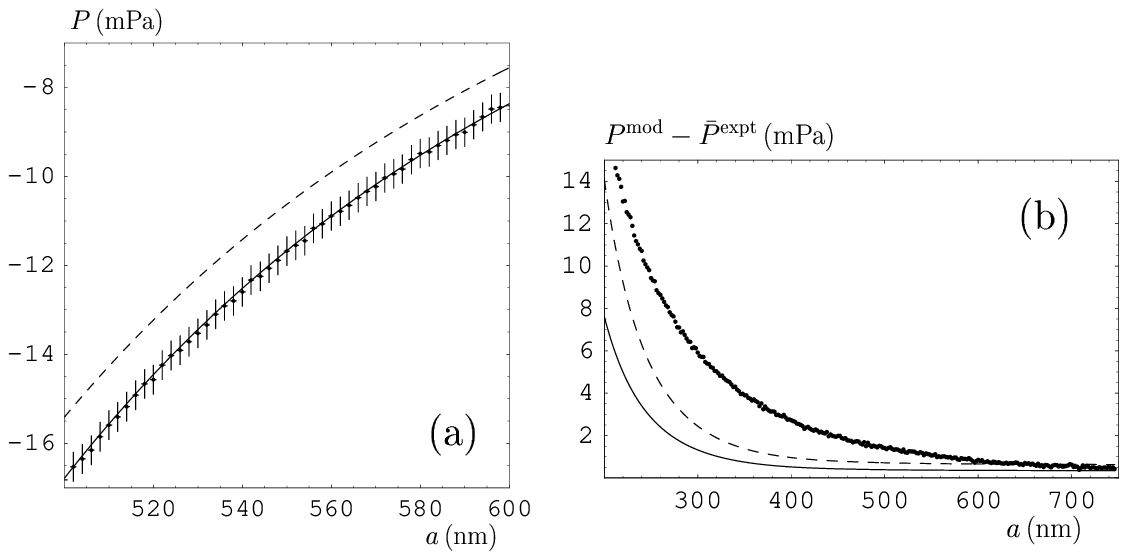}
\end{center}
\vspace*{-21.cm}
\caption{(a) The crosses show the measured mean Casimir pressures
together with the absolute errors as a function of the separation.
The theoretical Casimir pressures computed using the generalized
plasma-like model and the approach including the screening effects are
shown as solid and dashed lines, respectively.
(b) Differences of the
theoretical Casimir pressures computed with inclusion of the screening
effects and the  mean experimental Casimir pressures versus separation
are shown as dots. The 95\% and 99.9\% confidence intervals are shown
as the solid and dashed lines, respectively.}
\end{figure}
The solid line presents the computational results for
$P(a,T)=-\partial{\cal F}(a,T)/\partial a$ using the Lifshitz formula and
the generalized plasma-like dielectric permittivity (\ref{eq29}).
The parameters of oscillators for Au were determined in [5] with high
precision. The dashed line was computed using the Lifshitz formula for
$P^{\rm mod}(a,T)=-\partial\tilde{\cal F}(a,T)/\partial a$ with the
reflection coefficients $\tilde{r}_{\rm TM,TE}$ taking the
screening effects into account. As is seen in Fig.~6(a), the theoretical
approach taking into account the Thomas-Fermi screening length is
experimentally excluded at a 95\% confidence level over the separation
region from 500 to 600\,nm. The same conclusion follows within the
entire measurement range from 160 to 750\,nm.

Fig.~6(a) illustrates the
first method for the comparison
between experiment and theory in Casimir force
measurements discussed in Sec.~5. In Fig.~6(b) the second method for
the comparison of experiment and theory is illustrated. Here, the
differences between the theoretical Casimir pressures computed with
inclusion of the screening effects and the mean experimental pressures
are shown as dots. The solid line indicates the borders of 95\%
confidence intervals. Dots are outside the confidence interval
$[-\Xi_{P}(a),\Xi_{P}(a)]$ over the entire measurement range from 160
to 750\,nm. In the same figure, the dashed line shows the borders of 99.9\%
confidence intervals. As is seen in Fig.~6(b), dots are outside of this
confidence interval within the separation region from 160 to 640\,nm.
Thus, within this region of separations the theoretical approach taking
the screening effects into account [66] is experimentally excluded at
a 99.9\% confidence level.

The physical reasons why the inclusion of the screening effects
into the Lifshitz theory is thermodynamically and experimentally
inconsistent can be understood as follows. The Lifshitz theory is formulated
for systems in thermal equilibrium. As was indicated in [71], the drift
current of conduction electrons leads to heating of the crystal lattice.
In this case, if the constant temperature is preserved, there must be
a unidirectional flux of heat from the Casimir plates to the heat reservoir.
The existence of such an interaction between a system and a heat reservoir
is strictly prohibited in a state of thermal equilibrium [72] and is in
contradiction with its definition [73]. According to this definition,
in thermal equilibrium all irreversible processes connected with the
dissipation of energy are terminated. Specifically, in thermal equilibrium
any nonzero gradients of charge carrier density and any diffusion
are impossible. Thus, the inclusion of the screening effects and diffusion
currents into the Lifshitz theory is in violation of its applicability
conditions.

\section{Conclusions and discussion}

In the above, we have discussed several problems at the interface between
field-theoretical description of the Casimir effect and experiments on
measuring the Casimir force. The consideration of the Casimir energies
and forces in ideal metal rectangular boxes leads to the conclusion
that even when using ideal models it is important to take into account
some general physical requirements. Thus, it is not productive to use
the free energy which leads to the Casimir forces of quantum nature
which increase with increasing size of the box. It also seems
thermodynamically inconsistent to claim that the Casimir force acting
on a piston is a well defined quantity, whereas the forces acting on all
other faces of the box are excluded from consideration. The reason is
that if the free energy is defined correctly (see Sec.~2), there is a
 uniquely defined pressure on all faces of the box equal to the negative
derivative of the free energy with respect to the box volume calculated
at constant temperature.

An important tool for the comparison of experiment with theory is the
proximity force approximation. In Sec.~3 we have discussed some
inexact formulations which can be found in theoretical publications on
this subject. We have also discussed recent achievements in
quantum-field-theoretical approach to the calculation of the Casimir
energies in terms of functional determinants and scattering matrices.
This scientific direction has already obtained the first analytical results
beyond the PFA. It is of great promise for many experimentally relevant
applications of the theory.

In Secs.~4 and 5 we tried to add clarity to the widely discussed problems
of the precision of experiments on the Casimir effect and the
agreement between experiment and theory. It was stressed that the precision
of some independent measurements can be much higher than
of respective theoretical computations using the values of parameters
which may not be known precisely enough. In such cases the agreement
of experiment with theory can also be not as good as the precision of the
measurements.

Finally, in Sec.~6 we have analyzed a recent theoretical approach to the
thermal Casimir force taking into account the screening effects and
diffusion currents. Using quantum Fermi-Dirac statistics and respective
Thomas-Fermi screening length, we have applied this approach to
calculate the Casimir free energy between two metallic plates. It was
shown that the obtained free energy results in a violation of the Nernst
heat theorem for metals with perfect crystal lattices. Thus, the approach
under consideration is inconsistent with thermodynamics. The calculational
results for the Casimir pressure in the configuration of two Au plates
were compared with the results of the most precise experiment performed using
a micromachined oscillator. It was shown that the theoretical predictions
following from the inclusion of the screening effects are rejected by the
experimental data at a 99.9\% confidence level. The reason for the failure
of this approach is the inclusion of irreversible diffusion processes
violating thermal equilibrium which is the basic applicability condition
of the Lifshitz theory.

Phenomenologically, the Lifshitz theory combined with the generalized
plasma-like dielectric permittivity provides a description of dispersion
forces between metallic  test bodies which is in agreement with
thermodynamics and consistent with all available experimental information.
For now there is no other theoretical approach to the thermal Casimir
force between  metals
which would satisfy the requirements of thermodynamics and be
simultaneously consistent with all measurement data.

\ack{This work was supported by Deutsche Forschungsgemeinschaft,
 Grant No 436 RUS 113/789/0--4.
The author is grateful to the Center of Theoretical Studies and Institute
of Theoretical Physics, Leipzig University where this work was
performed for kind hospitality.}

\smallskip

\end{document}